# Unifying Multimodal Source and Propagation Graph for Rumour Detection on Social Media with Missing Features

Tsun-Hin Cheung, Kin-Man Lam

Department of Electronic and Information Engineering, The Hong Kong Polytechnic University, Kowloon, Hong Kong SAR, China

Email addresses: tsun-hin.cheung@connect.polyu.hk (T. Cheung), enkmlam@polyu.edu.hk (K. Lam)

**Abstract**

With the rapid development of online social media platforms, the spread of rumours has become a critical societal concern. Current methods for rumour detection can be categorized into image-text pair classification and source-reply graph classification. In this paper, we propose a novel approach that combines multimodal source and propagation graph features for rumour classification. We introduce the Unified Multimodal Graph Transformer Network (UMGTN) which integrates Transformer encoders to fuse these features. Given that not every message in social media is associated with an image and community responses in propagation graphs do not immediately follow source messages, our aim is to build a network architecture that handles missing features such as images or replies. To enhance the model's robustness to data with missing features, we adopt a multitask learning framework that simultaneously learns representations between samples with complete and missing features. We evaluate our proposed method on four real-world datasets, augmenting them by recovering images and replies from Twitter and Weibo. Experimental results demonstrate that our UMGTN with multitask learning achieves state-of-the-art performance, improving F1-score by 1.0% to 4.0%, while maintaining detection robustness to missing features within 2% accuracy and F1-score compared to models trained without the multitask learning framework. We have made our models and datasets publicly available at: https://thcheung.github.io/umgtn/.

1. Introduction

Social media has become a primary platform for individuals to consume online news, but it also facilitates the rapid spread of unverified or false information, giving rise to significant social concerns. Rumours, which consist of unconfirmed or misleading information, pose challenges for the public and even professional journalists in discerning truth amidst breaking news [1]. Therefore, the automatic detection of rumours on social media is of utmost importance.

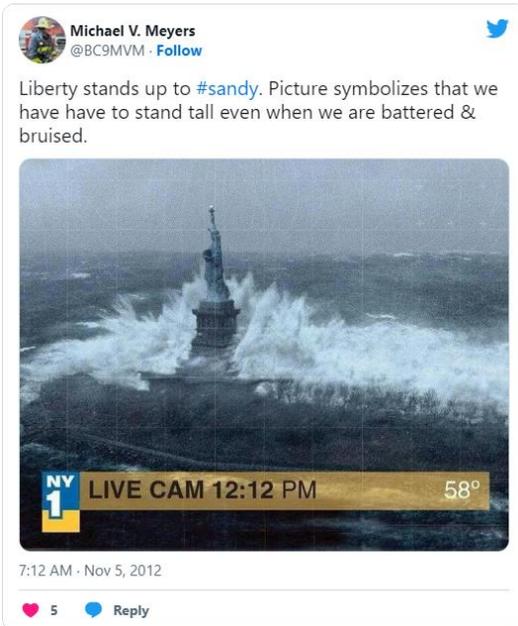 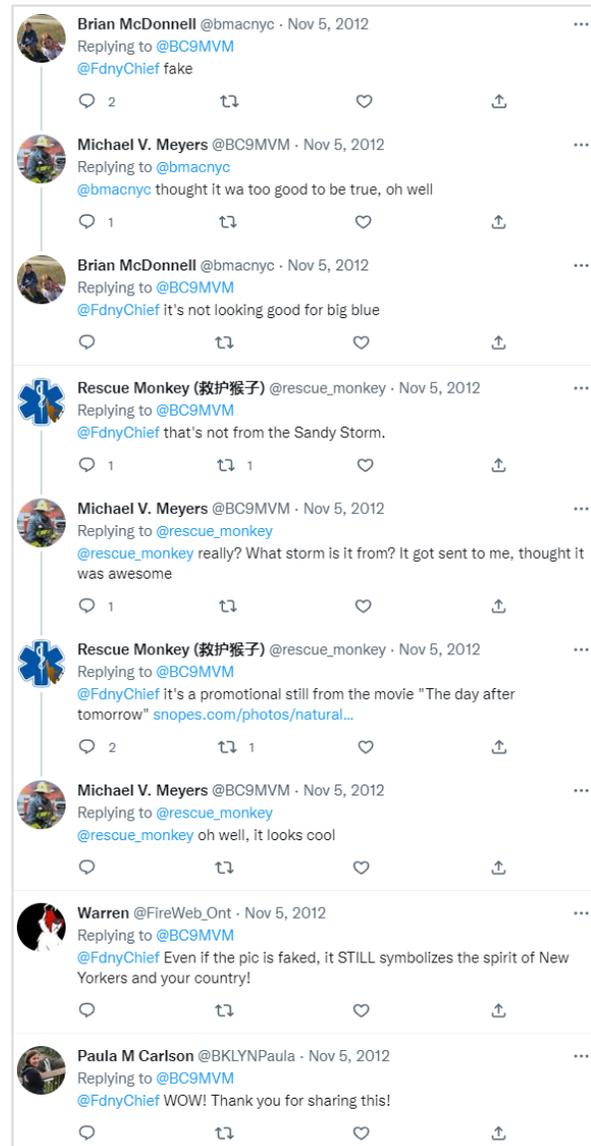

(a) (b)

**Fig. 1.** Example of (a) a multimodal source and (b) a propagation graph, i.e., replies, of a rumour on social media.

Contemporary online news often comprises multimodal content, incorporating both text and images. To effectively identify rumours on social media, it is crucial to consider both the visual and textual features of source messages and their corresponding replies in conversations. Multimodal rumours often involve forged or computer-generated imagery alongside textual descriptions. For instance, Fig. 1(a) presents an example of a multimodal rumour, where a misleading message is accompanied by a fake picture falsely claiming to depict Hurricane Sandy in 2012, which circulated on social media. The attached image is computer-generated and extracted from a movie. The combination of textual and visual features serves as vital cues for accurate discrimination between rumours and genuine news. Furthermore, the replies in social media conversations often contain opinions and judgments regarding the veracity of the source information. Fig. 1(b) illustrates replies that include phrases such as "really?" and "fake," expressing doubt and disagreement with the source information. These linguistic features serve as crucial indicators for learning models to classify rumours, especially within the context of social media. Thus, it is imperative to leverage both the multimodal source and propagation graph, i.e., the replies to the source, concurrently for rumour detection on social media.

Existing methods for detecting rumours on social media can be broadly categorized into two groups: image-text pair classification and source-reply graph classification. Image-text pair classification involves classifying pairs consisting of the text and image in the source information, considering both intermodal and intramodal relationships. Deep neural networks have been extensively employed to extract and fuse visual and textual features for multimodal rumour detection on social media. Previous studies commonly employed Convolutional Neural Networks (CNNs) for visual feature extraction and Recurrent Neural Networks (RNNs) for textual feature extraction [2], [3]. Recently, Bidirectional Encoder Representations from Transformers (BERT) [4] have been applied to extract textual features, yielding significant performance improvements in multimodal rumour detection when combined with attention mechanisms [5]–[7]. However, these studies overlook the valuable information present in the community responses to the source information, which has proven effective for rumour detection.

In the realm of source-reply graph classification, the propagation graph of community responses plays a crucial role. Existing approaches aim to capture the spatial and temporal relationships among messages within a conversation. Spatial relationships refer to the semantic correlation between a message and its linked replies in a propagation graph while temporal relationships denote the semantic evolution among all messages over time. To model the graph propagation pattern of social media conversations [8], [9]. Graph Neural Networks (GNNs) [10] are commonly utilized to model the graph propagation patterns in social media conversations by learning the spatial relationship between a message and its linked replies. Additionally, Recurrent Neural Networks (RNNs) [11] and Transformers [12] are employed to extract the temporal information within conversations. Hybrid models [13] have recently been proposed to represent dynamic graphs using a combination of GNNs and Transformers for detecting fake news on social media. Although separate studies have extensively investigated multimodal source-based and propagation graph-based rumour detection methods, the fusion of these two crucial features has received limited attention. This motivates our research to explore the co-occurrence of multimodal source and propagation graph features for rumour detection using a unified learning model and shared benchmark datasets.

Our study tackles two major challenges in rumour detection. First, we need to effectively fuse multimodal sources with propagation graph features to accurately identify rumours on social media. To achieve this, we propose the Unified Multimodal Graph Transformer Network (UMGTN), which leverages the Multimodal Transformer and Graph Transformer networks to learn and fuse multimodal source and propagation graph features using attention mechanisms.

Second, on social media, not every message is accompanied by images, and community responses do not immediately follow the posting of a source claim. Thus, our second challenge is to develop a learning model that is flexible in handling data with missing features during both training and inference. To address this, we employ multimodal Transformer blocks with a masking mechanism [14], allowing the network to learn from inputs containing missing features. Furthermore, we utilize a multitask learning framework [15] to process data with missing features, such as missing images or replies. This approach enhances the robustness and flexibility of our proposed model for rumour detection on social media.

The main contributions of our work are as follows:

- We propose the fusion of multimodal source information, combining source posts with images and source-reply propagation graphs, for rumour detection on social media. To achieve this, we introduce the Unified Multimodal Graph Transformer Network (UMGTN), which incorporates attention mechanisms to learn and fuse multimodal source and propagation graph features.
- We integrate an attention mask into the proposed UMGTN, allowing for the handling of feature-incomplete conversations from social media, where either images or replies may be missing. Moreover, we adopt a

- multitask learning framework to improve the robustness of our model to missing modalities, simultaneously training samples with complete and incomplete features.
- To evaluate the performance of our proposed method, we extend existing Twitter, PHEME, and Weibo datasets by collecting images and replies that were not included in the original datasets, using the released meta URLs. These extensions enable us to perform multimodal-graph rumour detection on social media.
- We conduct experiments on these real-world datasets to investigate the proposed network architecture and multitask training framework for rumour detection. The experimental results demonstrate the effectiveness and robustness of our proposed method across English and Chinese datasets.

## 2. Related Work

Detecting rumours and misinformation has been an active research area in the past decade. In this section, we provide an overview of related work on rumour detection, including text-based, multimodality-based, and propagation-based rumour classification. Additionally, we discuss the methods for handling missing features and the use of Transformers, which are relevant to our multitask training strategy for rumour detection.

*2.1. Text-based Classification*

Conventional rumour detection approaches are based on the statistical analysis of linguistic features. Takahashi et al. [16] were the first to analyse the spread of rumours on Twitter by considering word distributions in datasets. However, the analysing targets are limited to certain events, making it difficult to generalize the models to unseen events. Machine learning-based methods have shown promising performance in detecting rumours by utilizing linguistic features. Kwon et al. [17] were the first to employ machine learning-based methods, including decision trees, random forests, and Support Vector Machines (SVM), with linguistic features for rumour detection. However, these handcrafted features result in limited effectiveness and generalization ability.

Neural networks, such as Recurrent Neural Networks (RNNs) and Convolutional Neural Networks (CNNs), can effectively discover and learn hidden patterns in text data. These models are often used in conjunction with pretrained word embeddings, such as GloVe [18], to represent the semantic meaning of each word. Inspired by neural translation models, the Bidirectional Encoder Representation from Transformers (BERT) [4] was proposed for text classification. BERT consists of self-attention modules and represents each word as a linear combination of other word representations in a sentence. It has been widely used to encode sentences into fixed-length representations [19]. In our work, we leverage a pretrained Transformer for message representation, as it has been proven effective at encoding sentences for various text classification tasks, including fake news and rumour classification on social media [20].

*2.2. Multimodality-based Rumour Classification*

Text classification has been the most popular technique for social media analysis. However, with the increasing diversity of social media content, computer vision and image processing techniques have become valuable for analysis. Multimodal learning, which combines multiple modalities of information, has been applied to social media analysis, including fake news and rumour detection. Deep neural networks, capable of automatically learning deep representations for multimodal classification, have been commonly used to combine textual and visual information. For instance, Wang et al. [3] proposed an event-invariant adversarial neural network that learns multimodal domain-invariant features of a source post by using adversarial neural networks to remove event-specific features. Khattar et al. utilized a variational autoencoder, jointly learned with rumour classification, to extract a shared multimodal representation from textual and visual features. These approaches have shown improved performance in multimodal rumour classification. Recently, pretrained BERT has been adopted to replace RNNs for encoding text messages, significantly boosting the performance of multimodal rumour classification [22].

However, these methods often adopt simple fusion techniques, such as concatenation, to combine textual and visual representations, which may not fully exploit the intramodal relationship between the two modalities. To address this limitation, attention mechanisms have been applied to extract the deep correlation between text and image, resulting in more accurate rumour detection. For example, Jin et al. [23] proposed a multimodal recurrent neural network that combines visual and textual features using an attention mechanism for rumour detection. Zhou et al. [24] utilized image captioning with the LSTM network to explicitly learn the similarity and dissimilarity between the source text and image for fake news detection. Ying et al. [5] employed BERT with a cross-attention network to fuse visual and textual representations, forming a robust multimodal representation for misinformation detection. Chen et al. [25] proposed an ambiguity learning module that models the correlated and complementary relationship between textual and visual information. Furthermore, some researchers have considered the interactions between the source and replies from different people, which can enhance the accuracy and robustness of detection [26].

*2.3. Propagation-based Rumour Classification*

To improve the robustness of rumour detection, the propagation graph of a conversation, including the replies to the source information, is considered. Zubiaga et al. [27] constructed the PHEME dataset and developed logistic regression with conditional random fields (CRF) for rumour detection, using linguistic features of the source and replies for classification. Ma et al. [28] constructed the first Chinese rumour dataset by collecting sources and replies of real and fake news from Weibo. They proposed to adopt tree-based recursive neural networks (RvNNs) to model the time-series linguistic features from the source and replies for rumour classification. Subsequently, different neural network architectures were explored for source-reply graph classification. These methods typically consider spatial and temporal features in the replies, where spatial features represent the semantic dependence between a message and its replies, while temporal features refer to the sequential relationship among all the replies in a time-series manner [9]. These two features have been proven effective for rumour detection on social media.

To model the spatial relationship between a message and its replies, CNNs and GNNs are commonly used to extract features from the replies. For instance, Yu et al. [29] proposed a CNN-based network that utilizes convolutional kernels to learn the spatial relationship among the replies by grouping relevant posts as a fixed-length representation. Bian et al. [10] employed a graph convolutional network to learn the propagation patterns of the source and replies and utilized convolutional kernels to learn the relationship between the replies. In temporal-based methods, RNNs and Transformers have been widely studied. Ma et al. [11] proposed an RNN to learn the long-term dependence among the replies to the source information by considering the replies as a variable-length time series of responses. The method was further improved by using Transformers [30] to enhance the temporal representation of the source-replies graph. Vu et al. [12] integrated spatial and temporal features by using GNN to extract spatial features in a propagation graph, followed by an RNN to aggregate the flattened node features generated by GNN. Song et al. [13] used the Temporal Graph Network [31] to incorporate temporal information into a graph attention network, generating a comprehensive representation graph for source-reply graph classification. However, most of the propagation graph-based methods ignore the multimodal features in source posts, which are particularly useful for early rumour detection, as the propagation graph cannot be generated immediately when a source message is posted on social media. Therefore, this motivates us to study the co-occurrence of visual and propagation features for rumour detection.

*2.4. Classifying Data with Missing Features*

Learning to classify missing features is a common challenge in solving real-world machine learning problems [32]. In social media, a user may or may not upload images together with text in a source claim. Although Transformer-based neural networks have been proven effective in both multimodal source and source-reply graph rumour detection, the robustness to data with missing features in rumour detection is rarely studied. Ma et al. [15] proposed a multitask learning framework based on the Vision and Language Transformer [33], to improve the generalization ability of unimodal classification in a bimodal Transformer model. The multimodal Transformer is simultaneously trained with unimodal and bimodal data, which can alleviate the data missing problem and improve the detection performance when images are absent.

3. **Unified Multimodal Graph Transformer Network (UMGTN)**

In this section, we present our proposed Unified Multimodal Graph Transformer Network (UMGTN) for rumor detection. Fig. 2 provides an overview of the proposed model. The input of UMGTN consists of a multimodal source post and a source-reply graph. The multimodal source post includes a source text $S_T$ and a source image $S_I$. The source-reply graph consists of a node feature matrix $F$, obtained through the message embedding module, and an edge connectivity matrix $C$, that indicates the links to each reply. The goal is to classify a multimodal conversation into two categories: {*Rumour*, *Non-Rumour*}.

The proposed UMGTN incorporates three fusion modules: Multimodal Transformer Network (MTN), Graph Transformer Network (GTN), and Multimodal Graph Transformer (MGT). MTN models the intermodal correlation between the source text and image. GTN extracts spatial features, representing the semantic relationship between a post and its linked replies, within a graphical structure. MGT fuses temporal relationship, capturing the semantic relationship between a multimodal source feature and all its replies in a time-series manner, to form the final conversation representation.

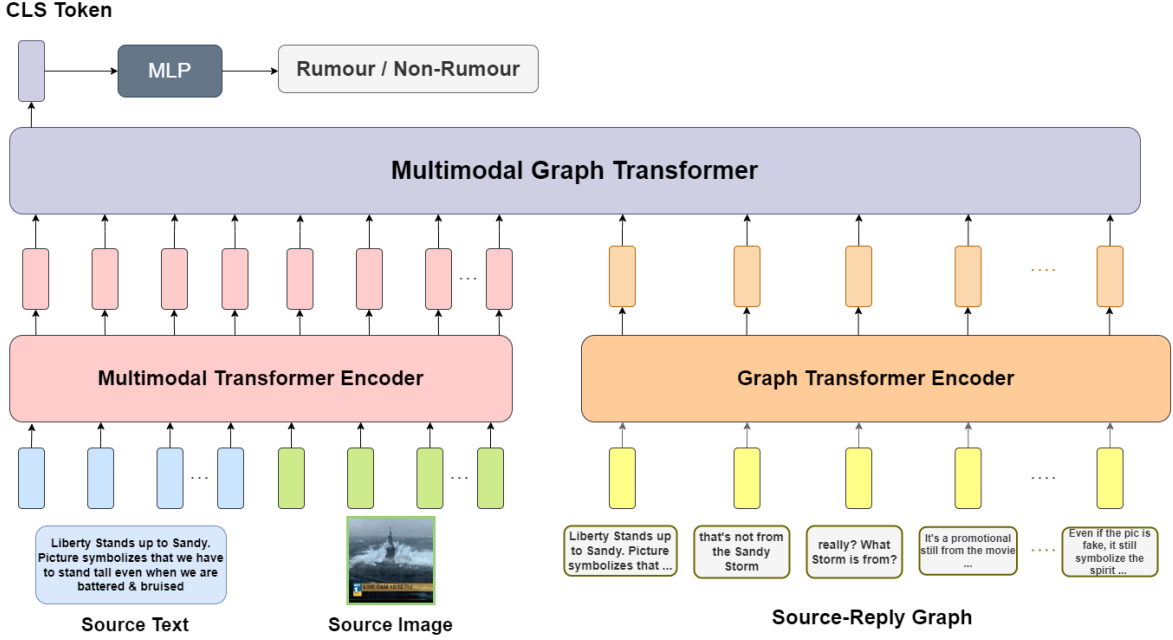

**Fig. 2.** The Proposed Unified Multimodal Graph Transformer Network (UMGTN)

*3.1. Transformer Encoder*

We introduce the standard Transformer encoder [34] and its attention masking mechanism, which serve as building blocks for MTN, GTN, and MGT. We adopt a standard implementation of Transformer without modifications to ensure easy implementation and reproducibility.

The Transformer model computes attention scores between all vectors in a sequence, enabling the learning of sequence representations. It consists of multiple multi-head attention modules. Each multi-head attention module consists of scaled dot-product attention blocks. These blocks accept three inputs: query $\mathcal{Q}$, key $\mathcal{K}$, and value $\mathcal{V}$. The attention weights are computed by measuring the similarity between the query matrix $\mathcal{Q}$ and key matrix $\mathcal{K}$. The attention vectors are obtained by multiplying the attention weights with the value matrix $\mathcal{V}$ to obtain the attention vectors. Mathematically, the attention score matrix $\mathcal{S}$ is calculated as follows:

$$\mathcal{S} = \text{Softmax}\left(\frac{\mathcal{Q}\mathcal{K}^{\text{T}}}{\sqrt{d_k}}\right), \tag{1}$$

where $d_k$ is the embedding dimension of the key $\mathcal{K}$. After calculating the attention score $\mathcal{S}$, the attention value $\mathcal{V}'$ can be computed as a linear combination of the attention score $\mathcal{S}$ and the key $\mathcal{K}$, as follows:

$$\mathcal{V}' = \mathcal{S}\mathcal{V}. \tag{2}$$

The multi-head attention mechanism repeats the scaled dot-product attention $h$ times, and aggregates the attention value $\mathcal{V}'$ to obtain a more robust representation of the value $\mathcal{V}$, as follows:

$$\text{MultiHead}(\mathcal{Q}, \mathcal{K}, \mathcal{V}) = \text{Concat}(V'_1 W'_1, V'_2 W'_2, \dots, V'_h W'_h) W^o, \tag{3}$$

where $\text{MultiHead}(\ )$ and $\text{Concat}(\ )$ represent the multi-head attention mechanism and the concatenation operation, respectively. It is worth noting that $W^o$ and $W'_i$ are trainable parameters, jointly learned through backpropagation. In the self-attention mechanism, we set the Query $\mathcal{Q}$, Key $\mathcal{K}$ and Value $\mathcal{V}$, equal to input $X$.

$$\text{Transformer}(X) = \text{LayerNorm}(\text{MultiHead}(\mathcal{Q}, \mathcal{K}, \mathcal{V}) + \mathcal{V}). \tag{4}$$

An attention mask is used to mask the query vector with respect to the key vector, thereby forcing the query not to pay attention to the key vectors at the masked positions. It is particularly useful to mask the padding positions of variable-length sequences in mini-batch processing. In our work, we utilize attention masks not only to mask missing features, but also in graph structure learning in GTN. A visualization of attention masks is shown in Fig. 3.

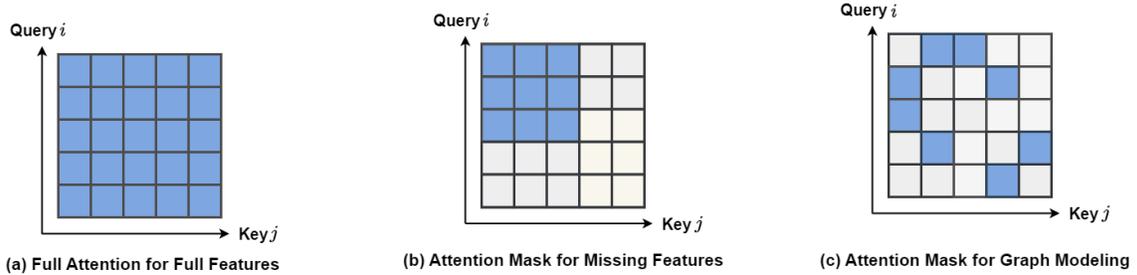

**Fig. 3.** Visualization of Attention Masks. (a) Mask for Full Features, (b) Mask for Missing Features. (c) Mask for Graph Modelling. The grey region in Row $i$ and Column $j$ represents masking the attention weights calculated from query vector $i$ and key vector $j$.

*3.2. Multimodal Transformer Network (MTN)*

To learn the semantic relationship between a source text and an image, we first use a pretrained BERT [4] and Vision Transformer (ViT) [35] to extract features from text and image, respectively. Specifically, the extracted features of the source text are the last hidden representations of the textual tokens $T = \{t_1, t_2, \dots, t_n\}$, where $n$ is the number of tokens in a source text. Similarly, we represent an image with the hidden representations of the visual tokens $V = \{v_1, v_2, \dots, v_m\}$, where $m$ is the number of tokens in a source image. Then, we apply two single linear layers, each denoted as $FC$, to project the textual and visual features, respectively, into a shared embedding space. After the linear projection layers, we concatenate the textual and visual features, which are then sent to a Transformer Encoder, as follows:

$$\boldsymbol{T'} = FC(\boldsymbol{T}), \tag{5}$$

$$\boldsymbol{V'} = FC(\boldsymbol{V}), \tag{6}$$

$$\boldsymbol{S} = Transformer([\boldsymbol{T'}, \boldsymbol{V'}]), \tag{7}$$

where [ ] is a concatenate operation between two sequences. It is worth noting that not all conversations are associated with images. We apply an attention mask, which is a tensor filled with zeros, so that our multimodal model can still be trained, when a conversation has no accompanied images, as shown in Fig. 3(b). We denote the output of MTN as $\boldsymbol{S}$.

*3.3. Graph Transformer Network (GTN)*

To obtain the node feature matrix $\boldsymbol{F}$ from a conversation, we first transform each message into a fixed-length representation. Each message is encoded by a pretrained BERT into a d-dimensional vector. Specifically, we use the hidden representation of the first token, i.e., a special learnable embedding vector added at the beginning of every sentence before sending it to the pretrained BERT, to form an overall representation of a sentence. For each message $\boldsymbol{m}_i$, we encode it as an embedding vector $\boldsymbol{f}_i \in \mathbb{R}^d$. After processing the source post and the replies with sentence embeddings, the node features $\boldsymbol{F} = \{\boldsymbol{f}_s, \boldsymbol{f}_1, \boldsymbol{f}_2, ..., \boldsymbol{f}_{k-1}\} \in \mathbb{R}^{d \times k}$ are obtained, where the first element in $\boldsymbol{F}$, i.e., $\boldsymbol{f}_s$, is the feature representation of the source information, and the other elements are that of the replies, i.e., $k-1$ replies.

Given a source tweet and its replies, we use a graph attention network [36] to learn the spatial features between a message and its linked replies. The input of the graph attention network is the node feature matrix $\boldsymbol{F}$ and the edge connectivity matrix $\boldsymbol{C}$. Our goal is to generate a more comprehensive feature for each node, by considering the relationship between the node and its neighbouring nodes, as follows:

$$\boldsymbol{F'} = FC(\boldsymbol{F}), \tag{8}$$

$$\boldsymbol{G} = Transformer(\boldsymbol{F'}). \tag{9}$$

We construct an attention mask using the connectivity matrix $\boldsymbol{C}$, so that messages are only allowed to reach those masked nodes with an edge connected as shown in Fig. 3(c). The output of GTN is denoted as $\boldsymbol{G}$.

*3.4. Multimodal Graph Transformer (MGT)*

Having obtained the multimodal source $\boldsymbol{S}$ and propagation graph features $\boldsymbol{G}$, we use another multimodal Transformer layer to integrate these two feature sequences, because the multimodal Transformer encoder allows every token to pay attention to all other tokens in a multimodal sequence in the temporal dimension [37]. Similar to the MGT module, we apply linear projection layers to project the multimodal source $S$ and the propagation graph features $G$ in a shared embedding space. After the linear projection layers, we concatenate the multimodal source and propagation graph features, which are then sent to MGT, as follows:

$$\boldsymbol{S'} = FC(\boldsymbol{S}), \tag{10}$$

$$\boldsymbol{G'} = FC(\boldsymbol{G}), \tag{11}$$

$$\boldsymbol{H} = Transformer([\boldsymbol{S'}, \boldsymbol{G'}]). \tag{12}$$

We use the first element of $\boldsymbol{H}$, i.e., the classification token, denoted as $\boldsymbol{h}_{cls}$, as the final representation of a multimodal conversation, which is forwarded to the classification layer for rumour detection.

*3.5. Classification and Loss Function*

Having obtained the feature representation $\boldsymbol{h}_{cls}$ of a conversation, we use the Softmax classifier to predict whether it is a rumour or not, as follows:

$$\hat{\boldsymbol{y}} = \text{softmax}(\boldsymbol{W}_c \boldsymbol{h}_{cls} + \boldsymbol{b}_c), \tag{13}$$

where $\boldsymbol{W}_c \in \mathbb{R}^{2 \times d}$ and $\boldsymbol{b}_c \in \mathbb{R}^2$ are trainable parameters. We employ the cross-entropy loss as the objective function in our proposed method. Given the predicted label $\hat{\boldsymbol{y}}$ and the ground-truth label $\boldsymbol{y}$, the negative log-likelihood is minimized. Thus, we have

$$\text{loss} = -\big(\boldsymbol{y}\log(\hat{\boldsymbol{y}}) + (\boldsymbol{1} - \boldsymbol{y})\log(1 - \hat{\boldsymbol{y}})\big). \tag{14}$$

*3.6. Multitask Learning*

To improve the robustness of rumour detection with missing features, we apply multitask learning. Our goal is to train a neural network robust to missing images or replies. We simultaneously train our deep model on data with full and missing features, with a multitask loss. An illustration of multitask learning on data with missing features is shown in Fig. 4.

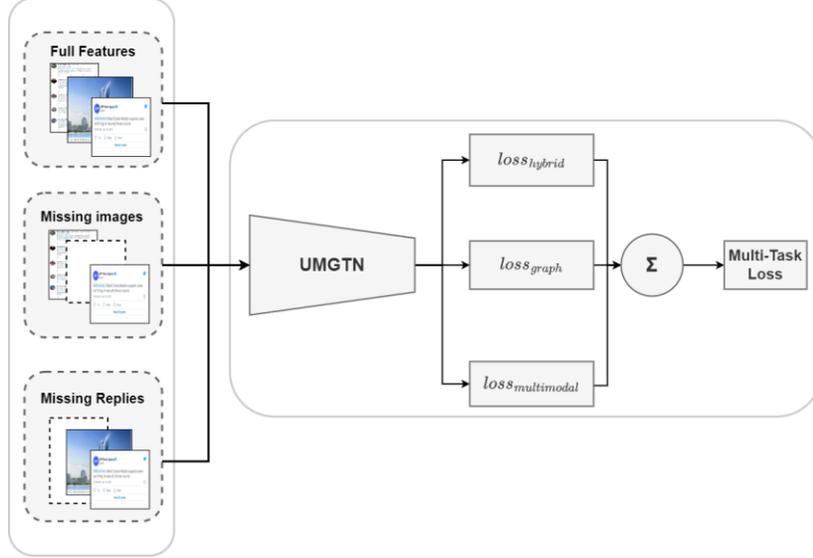

**Fig. 4.** Illustration of Multitask Learning using Data with Missing Features.

We compute the multitask learning loss as follows:

$$\text{loss}_{Total} = \text{loss}_{multimodal} + \text{loss}_{graph} + \text{loss}_{hybrid}, \qquad (15)$$

where $\text{loss}_{multimodal}$, $\text{loss}_{graph}$, and $\text{loss}_{hybird}$ represent multimodal classification, graph classification, and hybrid classification, respectively.

## 4. Experiment and Results

In this section, we first describe the datasets used and the experimental setup. Then, we evaluate the proposed model and compare its performance with existing state-of-the-art methods. After that, we present an ablation study of our proposed method for multimodal graph-based rumour detection. Finally, we visualise some rumours and non-rumours that can be classified by the proposed model.

*4.1. Datasets*

In the experiments, the PHEME [27], Weibo-16 [11], Twitter [38], and Weibo-17 [23] datasets were used to evaluate the methods for rumour detection. The PHEME and Weibo-16 datasets are widely used for graph-based rumour detection, while the Twitter and Weibo-17 datasets are widely used for multimodality-based rumour detection. Since all these datasets contain the URLs of the microblog posts from Twitter and Weibo websites, we can retrieve associated images and replies, which were not included in the original datasets, from the official Twitter[1] and Weibo[2] APIs. The detailed statistics of the original and recovered images and comments are shown in Table I.

| Datasets | PHEME [27] | Weibo-16 [11] | Twitter [38] | Weibo-17 [23] |
|---|---|---|---|---|
| # of Rumours claims | 1,972 | 2,313 | 6,226 | 4,748 |
| # of Non-Rumours claims | 3,830 | 2,351 | 9,405 | 4,779 |
| # of Total claims | 5,802 | 4,664 | 15,631 | 9,527 |
| # of Images in the original dataset | 0 | 0 | 410 | 9,527 |
| Total # of Available Images | 2,438 | 3,843 | 410 | 9,527 |
| # of Replies in the original dataset | 105,354 | 2,796,938 | 0 | 0 |
| Total # of Available Replies | 105,354 | 2,796,938 | 3,626 | 264,928 |

**Table I. Statistics of the datasets. The number of original images and replies for each dataset are released by the authors of the corresponding papers, while the recovered images and replies are combined with the original datasets and those collected by using our web API crawlers**.

**PHEME** [27]. This dataset contains 1972 rumour and 3830 non-rumour English conversations on Twitter, across five events, including Charlie Hebdo, Ferguson, Germanwings Crash, Ottawa Shooting, and Sydney Siege. It is found that two common experimental setups are adopted in most previous work. One is to perform leave-one-event-out cross-validation, so that there are no overlapping events between the training and testing sets. Another setup is to directly divide the dataset into training and testing sets with a ratio of 8:2. It is worth noting that the leave-one-event-out setting is closer to the real scenario, since rumours usually cover newly emerging topics on social media. However, to ensure fair comparisons, we conduct two experiments under the two different settings, and compare them with the corresponding models separately.

**Twitter** [38]**.** This benchmark dataset contains misleading multimedia content posted on Twitter. It is a commonly used dataset to evaluate the performance of multimodal rumour detection. The original dataset contains 410 images, shared by 15631 tweets. For a fair comparison to benchmarks, we filter out those tweets with videos attached. The number of non-rumour claims is 7557, while the number of rumour claims is 6208. This dataset has already been split into training and testing sets, ensuring that there are no overlapping events between the training and testing sets.

---

[1] Twitter API https://developer.twitter.com/en/docs/twitter-api
[2] Weibo API: https://open.weibo.com/wiki/API/en

**Weibo-16** [11]**.** The Weibo-16 dataset contains 2313 rumour and 2350 non-rumour Chinese conversations on Weibo. It was originally used for propagation graph-based rumour detection. We further collect the images, attached to the claims, using the official Weibo API. To ensure a fair comparison to previous work, we divide the Weibo-16 dataset in a ratio of 8:2, for training and testing, respectively.

**Weibo-17** [23]**.** The Weibo-17 dataset contains 6226 rumours and 9405 non-rumours Chinese conversations on Weibo. It was originally used for multimodal source-based rumour detection. We further collect the comments attached to the claims using the official Weibo API. To ensure a fair comparison to previous work, we divide the Weibo-17 datasets in a ratio of 8:2, for training and testing, respectively.

*4.2. Experimental Setup*

**Evaluation Metrics**: To evaluate the performance of different classification models for the rumour and non-rumour categories, the evaluation metrics, Precision, Recall and F1 scores, are used. We also measure the accuracy and macro F1 score, i.e., the mean F1 scores of positive and negative classes, to evaluate the performance of different rumour detection methods.

**Hyperparameters**: We use the pretrained English and Chinese BERTs [4], based on the English [39] and Chinese [40] datasets, respectively. The embedding dimension of the two BERT models is both 768. To extract multimodal features, we use the pretrained ViT and BERT, as the visual and textual backbones in the multimodal Transformer network, with an embedding dimension of 768. For all the experiments, the two models were trained with a mini-batch size of 16 for 30 epochs. The Adam optimizer is used with a fixed learning rate of 0.00005. To avoid overfitting, we use the L2 regularization with a rate of 0.001 and a dropout rate of 0.1. Our models were implemented in PyTorch [41], with the add-on of Torch-Geometric [42] and Transformers [43], for graph learning and Transformer modules, respectively. All experiments were conducted on two GeForce RTX 2080 Ti GPUs.

*4.3. Comparison with Multimodality-based Methods*

We compare our proposed model, i.e., UMGTN, with the following state-of-the-art methods for multimodality-based rumour detection. The quantitative results of the different methods are shown in Table II.

**NeuralTalk** [44]. This method was originally used to generate natural sentences to describe images. It is used to fuse the representations of image and text by averaging the output of RNN at each timestep.

**VQA**[45]. Visual question answering (VQA) aims to answer questions about given images. To apply VQA to detect rumours, the elementwise multiplication between text and image is replaced by feature concatenation.

**Att-RNN** [23]. Att-RNN uses LSTM to extract both text and social context features and obtain a joint representation, which is then combined with visual features extracted from a pretrained deep CNN with attention.

**EANN**[3]. Text-CNN and pretrained VGG-19 are employed to extract textual and visual features. Then, the multi-modal features are fed to a rumour detector to predict whether a post is a rumour or not.

**MVAE** [21]. MVAE aims to learn a shared representation between textual and visual modalities to detect rumours. A variational autoencoder is leveraged to obtain a shared representation by reconstructing the input data from the sampled multimodal representation.

**BDANN** [22]. The multimodal feature extractor employs a pretrained BERT model to extract text features and a pretrained VGG-19 model to extract image features. The extracted features are then concatenated and fed to a detector to distinguish fake news.

**CAFE** [25]. A cross-modal ambiguity learning-based method was proposed for multimodal fake news detection. Different from previous work, CAFE is capable of adaptively aggregating discriminative cross-modal correlation features and unimodal features based on the inherent cross-modal ambiguity.

**MEAN** [46]. This network adopts a multimodal generator and a dual discriminator. The multimodal generator extracts latent discriminant feature representations of text and image modalities. A decoder is employed to reduce information loss during the generation process for each modality.

**BCMF** [47]. Bidirectional Cross-Modal Fusion (BCMF) comprehensively integrates textual and visual representations in a bi-directional manner, using attention mechanisms.

| Dataset | Method | Acc. | Rumour | | | Non-rumour | | |
|---|---|---|---|---|---|---|---|---|
| | | | Pre. | Recall | F1 | Pre. | Recall | F1 |
| Twitter | NeuralTalk [44] | 0.610 | 0.728 | 0.504 | 0.595 | 0.534 | 0.752 | 0.625 |
| | VQA [45] | 0.631 | 0.765 | 0.509 | 0.611 | 0.550 | 0.794 | 0.650 |
| | Att-RNN [23] | 0.682 | 0.780 | 0.615 | 0.689 | 0.603 | 0.770 | 0.676 |
| | EANN [3] | 0.648 | 0.810 | 0.498 | 0.617 | 0.584 | 0.759 | 0.660 |
| | MVAE [21] | 0.745 | 0.801 | 0.719 | 0.758 | 0.689 | 0.777 | 0.730 |
| | BDANN [22] | 0.830 | 0.810 | 0.630 | 0.710 | 0.830 | 0.930 | **0.880** |
| | CAFE [25] | 0.806 | 0.807 | 0.799 | 0.803 | 0.805 | 0.813 | 0.809 |
| | MEAN [46] | 0.780 | 0.690 | **0.840** | 0.760 | **0.870** | 0.740 | 0.800 |
| | BCMF [47] | 0.815 | **0.854** | 0.805 | **0.829** | 0.772 | 0.827 | 0.799 |
| | **UMGTN (full)** | **0.842** | 0.770 | 0.809 | 0.790 | 0.886 | 0.861 | 0.873 |
| Weibo-17 | NeuralTalk [44] | 0.726 | 0.794 | 0.613 | 0.692 | 0.684 | 0.840 | 0.754 |
| | VQA[45] | 0.736 | 0.797 | 0.634 | 0.706 | 0.695 | 0.838 | 0.760 |
| | Att-RNN [23] | 0.788 | 0.862 | 0.686 | 0.764 | 0.738 | 0.890 | 0.807 |
| | EANN [3] | 0.782 | 0.827 | 0.697 | 0.756 | 0.752 | 0.863 | 0.804 |
| | MVAE [21] | 0.824 | 0.854 | 0.769 | 0.809 | 0.802 | 0.875 | 0.837 |
| | BDANN [22] | 0.842 | 0.830 | 0.870 | 0.850 | 0.850 | 0.820 | 0.830 |
| | CAFE [25] | 0.840 | 0.855 | 0.830 | 0.842 | 0.825 | 0.851 | 0.837 |
| | MEAN [46] | 0.894 | 0.900 | 0.870 | 0.890 | 0.890 | 0.910 | 0.900 |
| | BCMF [47] | 0.907 | 0.927 | 0.889 | 0.908 | 0.887 | 0.925 | 0.906 |
| | **UMGTN (full)** | **0.955** | **0.971** | **0.937** | **0.954** | **0.939** | **0.973** | **0.956** |

**Table II. Performance of the Proposed UMGTN and Other Multimodal source-based Methods. All results are extracted from the corresponding papers.**

Table II shows that our proposed model outperforms other state-of-the-art methods, in terms of accuracy and F1 score, for rumour detection. On the Twitter dataset, our method achieves much better performance than EANN [3], MVAE [21], and BDANN [22]. This is because these methods simply fuse visual and textual features by concatenation. Compared with BCMF [47], the proposed method performs slightly better, in terms of Macro-F1. Both BCMF [47] and our network also apply attention mechanisms to fuse textual and visual features, so they outperform EANN [3], MVAE [21], and BDANN [22]. For the Weibo dataset, our method achieves the best performance, in terms of both accuracy and F1 score. This is because MTN in our method is likely more robust to the Chinese language than all other methods. More importantly, the proposed UMGTN utilizes the propagation features. Thus, our method significantly outperforms the existing multimodality-based methods, especially on the Weibo dataset.

*4.4. Comparison with Graph-based Methods*

We compare our proposed model, i.e., UMGTN, with the following state-of-the-art methods for graph-based rumour detection. The quantitative results of the different methods are shown in Table III.

**DT-Rank** [48]. A decision tree is employed to detect rumours using rankings of enquiry phrases and clustering claims, with statistical features. This method was reimplemented and reported in [11].

**SVM-TS** [49]. A linear Support Vector Machine (SVM) is used to classify time-series structures, which model the handcrafted variation of propagation features. The method was implemented and reported in [11].

**GRU** [11]. A Gated-Recurrent Unit (GRU) is a type of Recurrent Neural Network (RNN) used to model the sequential information among messages from a conversation for rumour detection.

**CRF** [27]. A Conditional Random Field (SRF) model is proposed to detect rumours on the PHEME dataset. The authors used both content and social context-based features to train the model.

**ARN** [27]. This is an Attention-based Residual Network (ARN), which uses a CNN with residual connection to combine context features and social features with an attention mechanism, to find out the most important sets of features.

**GAN-GRU** [50]. This paper proposes a Generative Adversarial Network (GAN) based on GRU, to augment the training samples, by generating harder conversations, to force the discriminator to more robustly learn rumour-indicative features.

**CGAT** [51]. A Graph Adversarial learning (GAT) framework is proposed. The attacker network tries to dynamically add intentional perturbations on the graph structure to fool the detector, while the detector will learn more distinctive structural features.

**Bi-GCN** [10]. A Bi-directional Graph Convolutional Network (Bi-GCN) is proposed to explore both the characteristics by operating on the top-down and bottom-up propagation of rumours. It leverages a GCN with a top-down directed graph of rumour spreading to learn the pattern of rumour propagation.

**AARD** [52]. An adversary-aware adopted weighted-edge Transformer-Graph Network is proposed, with a position-aware adversarial response generator, to improve the vulnerability of detection models.

**ClaHi-GAT** [53]. A Claim-guided Hierarchical Graph Attention Network is used to augment the representation learning for responsive messages using the social contexts, and attends over the messages that can semantically infer the target claim.

**PPA-WAE** [54]. This network first models the propagation structure of each rumour as an independent set of propagation paths, in which each path represents the source post in a different talking context. Then, all paths are aggregated to obtain a representation of the whole propagation structure.

**Rumor2Vec** [55]. A novel rumour detection framework is used to jointly learn the text and propagation structure representation, by incorporating the propagation structures of all tweets to mitigate the sparsity issue.

**GACL** [51]. A Graph Adversarial Contrastive Learning (GACL) method is introduced by using the contrastive loss function to explicitly perceive the difference between conversational threads of the same class and different classes. Meanwhile, an Adversarial Feature Transformation (AFT) module is designed to produce conflicting samples for pressurizing the model to mine event-invariant features.

**RDCL** [56]. This is a contrastive learning framework for false information detection on social networks. It leverages contrastive learning to maximize the consistency between perturbed graphs from the same original graph and minimize the distance between perturbed and original graphs from the same class, thereby forcing the model to improve its resistance to data perturbations.

**CanarDeep** [57]. A hybrid deep neural model combines the predictions of a hierarchical attention network (HAN) and a multilayer perceptron (MLP) learned using social context-based (text and meta-features) and user-based features.

**DCNF** [58]. A dynamic propagation graph-based fake news detection method is proposed to capture the dynamic propagation information in static networks and classify fake news. Specifically, the proposed method models each news propagation graph as a series of graph snapshots recorded at discrete time steps.

**DA-GCN** [59]: An attentive graph neural network is proposed for rumour detection on social media. It aims to capture the most informative semantic and propagation features using dual-attention networks.

| Dataset | Method | Acc. | Rumour | | | Non-rumour | | |
|---|---|---|---|---|---|---|---|---|
| | | | Pre. | Recall | F1 | Pre. | Recall | F1 |
| PHEME (random-split) | DT-Rank [48] | 0.562 | 0.588 | 0.421 | 0.491 | 0.549 | 0.704 | 0.617 |
| | SVM-TS [49] | 0.651 | 0.663 | 0.617 | 0.639 | 0.642 | 0.686 | 0.663 |
| | GAN-GRU [50] | 0.781 | 0.773 | 0.796 | 0.784 | 0.791 | 0.766 | 0.778 |
| | AARD [52] | 0.848 | **0.863** | 0.829 | **0.845** | 0.837 | 0.868 | 0.852 |
| | ClaHi-GAT [53] | 0.859 | - | - | 0.790 | - | - | 0.893 |
| | GACL [51] | 0.850 | 0.801 | 0.750 | 0.772 | 0.871 | **0.901** | 0.885 |
| | RDCL [56] | 0.871 | - | - | 0.804 | - | - | 0.903 |
| | **UMGTN (full)** | **0.881** | 0.784 | **0.896** | 0.836 | **0.942** | 0.873 | **0.907** |
| PHEME (leave-one-event-out) | CRF [27] | 0.741 | 0.692 | 0.559 | 0.601 | 0.750 | 0.854 | 0.794 |
| | ARN [60] | 0.705 | 0.572 | 0.625 | 0.593 | 0.770 | 0.735 | 0.750 |
| | CanarDeep [57] | 0.725 | 0.668 | 0.603 | 0.631 | 0.738 | 0.784 | 0.758 |
| | **UMGTN (full)** | **0.780** | **0.675** | **0.714** | **0.691** | **0.822** | **0.794** | **0.807** |
| Weibo-16 | DT-Rank [48] | 0.732 | 0.738 | 0.715 | 0.726 | 0.726 | 0.749 | 0.737 |
| | SVM-TS [49] | 0.818 | 0.822 | 0.812 | 0.817 | 0.815 | 0.824 | 0.819 |
| | GRU [11] | 0.910 | 0.876 | 0.956 | 0.914 | 0.952 | 0.864 | 0.906 |
| | GAN-GRU [50] | 0.863 | 0.843 | 0.892 | 0.866 | 0.885 | 0.833 | 0.858 |
| | CGAT [61] | 0.940 | 0.959 | 0.906 | 0.932 | 0.925 | 0.968 | 0.946 |
| | Bi-GCN [10] | 0.961 | 0.961 | 0.964 | 0.961 | 0.962 | 0.962 | 0.960 |
| | Rumor2Vec [55] | 0.951 | 0.958 | 0.948 | 0.953 | 0.945 | 0.956 | 0.950 |
| | DA-GCN [59] | 0.944 | 0.941 | 0.946 | 0.944 | 0.947 | 0.941 | 0.944 |
| | PPA-WAE [54] | 0.962 | 0.963 | 0.965 | 0.964 | 0.961 | 0.960 | 0.961 |
| | DCNF [58] | 0.957 | 0.958 | 0.954 | 0.957 | 0.957 | 0.960 | 0.958 |
| | **UMGTN (full)** | **0.971** | **0.976** | **0.965** | **0.971** | **0.966** | **0.977** | **0.971** |

**Table III. Performance of the Proposed UMGTN and Other Graph-based Methods. The results of other methods are extracted from the corresponding papers. "-" means that the evaluation metric is not available in the corresponding paper.**

Compared to source-reply graph classification, our proposed method is much better than GRU [11], GAN-GRU [50], and Bi-GCN [10]. This is likely because these methods use either spatial-only or temporal-only features, which cannot provide robust representations for detecting rumours. Compared with recent state-of-the-art methods, including RDCL [56] and DCNF [58], which consider spatial and temporal features simultaneously, our method still achieves a slightly better result. This is because our proposed method uses both the multimodal source and source-reply graph features, which are essential for rumour classification. We will explore the importance of these two features in the ablation study of this paper.

*4.5. Ablation Study*

To verify the effectiveness of the proposed network, we conduct an ablation study on UMGTN. We remove the Multimodal Transformer Network, Graph Transformer Network, and Multimodal Graph Transformer, to examine the effectiveness of each module. The experimental results are shown in Table IV.

| Dataset | Module | F1 score (Positive) | F1 score (Negative) | Accuracy | F1 (Macro) |
|---|---|---|---|---|---|
| Weibo-16 | w/o Multimodal Transformer Network | **0.9708** | 0.9712 | **0.9710** | **0.9710** |
| | w/o Graph Transformer Network | 0.8790 | 0.8964 | 0.8884 | 0.8877 |
| | w/o Multimodal Graph Transformer | 0.9034 | 0.9097 | 0.9067 | 0.9066 |
| | UMGTN (full) | 0.9706 | **0.9714** | **0.9710** | **0.9710** |
| PHEME | w/o Multimodal Transformer Network | 0.8227 | 0.8917 | 0.8655 | 0.8572 |
| | w/o Graph Transformer Network | 0.8325 | 0.9057 | 0.8793 | 0.8691 |
| | w/o Multimodal Graph Transformer | 0.8293 | 0.9046 | 0.8776 | 0.8670 |
| | UMGTN (full) | **0.8365** | **0.9065** | **0.8810** | **0.8715** |
| Weibo-17 | w/o Multimodal Transformer Network | 0.9386 | 0.9425 | 0.9406 | 0.9406 |
| | w/o Graph Transformer Network | 0.9244 | 0.9293 | 0.9269 | 0.9269 |
| | w/o Multimodal Graph Transformer | 0.9174 | 0.9236 | 0.9206 | 0.9205 |
| | UMGTN (full) | **0.9538** | **0.9558** | **0.9548** | **0.9548** |
| Twitter | w/o Multimodal Transformer Network | 0.7843 | 0.8645 | 0.8335 | 0.8244 |
| | w/o Graph Transformer Network | 0.7882 | 0.8730 | 0.8412 | 0.8306 |
| | w/o Multimodal Graph Transformer | 0.7706 | 0.8687 | 0.8330 | 0.8197 |
| | UMGTN (full) | **0.7891** | **0.8734** | **0.8418** | **0.8313** |

**Table IV. Results of Ablation Study**

The experimental results show that the performance drops, when any one of the Graph Transformer Network, Multimodal Transformer Network, or Multimodal Graph Transformer is removed. The accuracy and F1 score are 2% to 4% higher on the PHEME, Twitter and two Weibo datasets, when all modules are used together. This is because the Transformer encoders can pay more attention to the important features of the multimodal source and propagation graph, while filtering out unnecessary features for rumour detection. Moreover, the importance of each of the three modules depends on the datasets. The performance drops the most when MTN is removed, when evaluated on the PHEME dataset. When MGN is removed, the performance drops the most when evaluated on the Weibo-17 and the

Twitter datasets. This is mainly because different datasets have different distributions of images and replies. Some of them contain highly correlated image-text pairs, while some contain important propagation graph features for identifying rumours. Nonetheless, our method can select important multimodal source and propagation graph features for rumour detection.

*4.6. Results of Multitask Learning*

For the multitask learning framework, we test the performance of multitask learning, by comparing the results of models trained with full-modal data. The qualitative results of the three models for hybrid classification, multimodal classification, and graph classification are shown in Table V.

| Dataset | Training Strategy | Task | F1 score (Positive) | F1 score (Negative) | Accuracy | F1 (Macro) |
|---|---|---|---|---|---|---|
| Weibo-16 | w/o Multitask Learning | Image-Text Source Classification | 0.7735 | 0.8300 | 0.8058 | 0.8018 |
| | | Source-Reply Graph Classification | 0.9529 | 0.9504 | 0.9517 | 0.9517 |
| | | Hybrid Classification | 0.9606 | 0.9600 | 0.9603 | 0.9603 |
| | with Multitask Learning | Image-Text Source Classification | 0.9034 | 0.9097 | 0.9067 | 0.9067 |
| | | Source-Reply Graph Classification | 0.9696 | 0.9703 | 0.9700 | 0.9700 |
| | | Hybrid Classification | **0.9706** | **0.9714** | **0.9710** | **0.9710** |
| Twitter | w/o Multitask Learning | Image-Text Source Classification | 0.7376 | 0.8488 | 0.8082 | 0.7932 |
| | | Source-Reply Graph Classification | 0.7336 | 0.8657 | 0.8214 | 0.8000 |
| | | Hybrid Classification | 0.7366 | 0.8485 | 0.8076 | 0.7926 |
| | with Multitask Learning | Image-Text Source Classification | 0.7706 | 0.8687 | 0.8330 | 0.8197 |
| | | Source-Reply Graph Classification | 0.7543 | 0.8700 | 0.8291 | 0.8122 |
| | | Hybrid Classification | **0.7891** | **0.8734** | **0.8418** | **0.8313** |
| Weibo-17 | w/o Multitask Learning | Image-Text Source Classification | 0.9255 | 0.9262 | 0.9259 | 0.9259 |
| | | Source-Reply Graph Classification | 0.9358 | 0.9400 | 0.9380 | 0.9379 |
| | | Hybrid Classification | 0.9496 | 0.9505 | 0.9501 | 0.9500 |
| | with Multitask Learning | Image-Text Source Classification | 0.9356 | 0.9392 | 0.9374 | 0.9374 |
| | | Source-Reply Graph Classification | 0.9463 | 0.9485 | 0.9474 | 0.9474 |
| | | Hybrid Classification | **0.9538** | **0.9558** | **0.9548** | **0.9548** |
| PHEME | w/o Multitask Learning | Image-Text Source Classification | 0.7718 | 0.8994 | 0.8603 | 0.8356 |
| | | Source-Reply Graph Classification | 0.8042 | 0.9054 | 0.8724 | 0.8548 |
| | | Hybrid Classification | 0.8123 | 0.9053 | 0.8741 | 0.8588 |
| | with Multitask Learning | Image-Text Source Classification | 0.8293 | 0.9046 | 0.8776 | 0.8670 |
| | | Source-Reply Graph Classification | 0.8331 | 0.9044 | 0.8784 | 0.8689 |
| | | Hybrid Classification | **0.8365** | **0.9065** | **0.8810** | **0.8715** |

**Table V. Results of UMGTN with or without Multitask Learning.**

The experimental results show that the overall performance using multitask learning is more robust to missing modalities, i.e., either the image or replies. For the Weibo dataset, without applying multitask learning, the F1 score for image-text classification drops by 12%. For the PHEME dataset, without applying multitask learning, the F1 score of source-reply graph classification drops by 4%. It is worth noting that the overall performance of hybrid classification is improved, when evaluated on all datasets. This is because the original datasets contain data with missing modality, i.e., posts without images or posts without replies. With the help of multitask learning, the generated modal-incomplete data can be viewed as data augmentation. Thus, the robustness of our model to missing features can be greatly improved by multitask learning. Furthermore, it is observed that propagation features are more important in the PHEME, Weibo-16 and Weibo-17 datasets, while the multimodal features are more important in the Twitter dataset. A reason for this is that the Twitter dataset contains a smaller number of replies compared to the other three datasets. It is worth noting that a community response may not appear immediately when a source message is posted on social media. With the multimodal source information, our model with multitask learning can perform relatively accurate rumour detection in the early stages of rumour propagation, which may prevent the spread of misinformation on social media as early as possible.

*4.7. Case Study*

In this section, we show some qualitative results for multimodal rumour detection, so as to understand the importance of visual and propagation features for rumour detection. We randomly selected some rumour conversations from both the Chinese and English datasets, which can be identified by the multimodal model. Fig. 5 shows the messages and their corresponding images.

In Fig. 5(a), we can observe that the source tweet is seeking donations due to an earthquake. This source is accompanied by a manipulated image, which shows a road cracked unrealistically. Although there are no replies to the source tweet, our method is still able to classify it as a rumour. In Fig. 5(b), the source image contains an unrealistically inclined tower. Our multimodal detection model can successfully classify it as a rumour, with the help of a combination of visual and textual information. Fig. 5(c) and Fig. 5(d) show some non-rumours posted on Twitter and Weibo, respectively. Fig. 5(c) contains an image of a man picking up a burning tear gas canister, while Fig. 5(d) contains an image of heavy rain and a flooded road. The two source texts are describing facts and the community responses, some of which support the source information, contain no judgments. Therefore, these two claims can be successfully classified as non-rumours by our method.

| | **Source Text** | **Image** | **Replies** | **Label** |
|---|---|---|---|---|
| (a) | Donate here to help the emergency response team of the #NepalEarthquake | 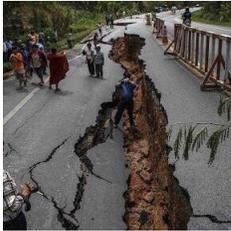 | Null | Rumour |
| (b) | 今天台湾地震把厦门一高层震歪了！台湾震中的楼没歪，厦门的楼被震歪了。<br>(Today's earthquake in Taiwan shook a high-rise building in Xiamen! The building at the epicenter of Taiwan was not crooked, but the building in Xiamen was crooked.) | 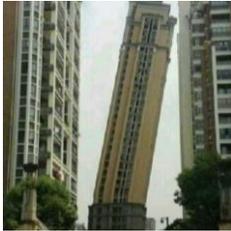 | (i) 肯定是 PS 的啊，大陆的楼能歪成这样还不倒！<br>(It must be Photoshop (PS), the buildings in the mainland can be crooked like this and still not fall down?)<br>(ii) 这真的假的啊<br>(Is real or fake?)<br>(iii) P 得这么假也有人信？<br>(The Photoshop technique is so fake, and some people believe it?)<br>(iv) 这图 p 的能再假点么<br>(Can this picture photo be faked?) | Rumour |
| (c) | RT @manofsteele: Man picks up burning tear gas can and throws it back at police. | 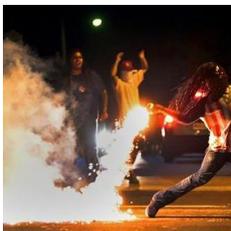 | (i) he's not spilling any chips or anything either.<br>(ii) ITH AN OPEN BAG OF CHIPS IS HIS HAND!<br>(iii) WITH AN OPEN BAG OF CHIPS IS HIS HAND! GREATEST PICTURE EVER. | Non-Rumour |
| (d) | 今早，暴雨、防汛防台均橙色预警，目前，青浦、普陀等多处已"见海"。其中，地铁二号线徐泾东路面积水严重。<br>(This morning, the rainstorm and flood control all woke up early. At present, Qingqing, Putuo and many other places have "see the sea". Among them, the water surface of No. 2 Xujing East Road is serious.) | 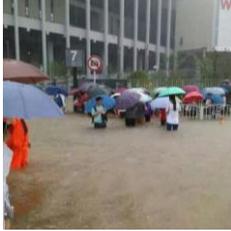 | (i) 出行注意安全…<br>(Pay attention to safety)<br>(ii) 简直暴雨好么？<br>(It's just raining)<br>(iii) 今天回家，积水太深，踮起脚尖过去了，鞋子湿透<br>(I went home today, the water was too deep, I tiptoed over, my shoes were soaked)<br>(iv) 我就没见过一次，浦东排水还不错<br>(I haven't seen it once, Pudong drainage is not bad) | Non-Rumour |

**Fig 5.** Visualization of correctly classified Rumours and Non-Rumours by UMGTN (with English Translation provided for Chinese conversations)

## 5. Implications of The Proposed Method

This section discusses the implications of the research work on rumour detection presented in this paper. The main objective of this work is to combine two different approaches, namely image-text source classification and source-reply graph classification, into a single model. While these approaches have been extensively studied individually, their fusion has received limited attention. Therefore, the proposed method aims to bridge this research gap by utilizing attention mechanisms to fuse propagation graph features and multimodal source information. Existing benchmark datasets for rumour detection typically focus on either multimodal-source or propagation graph features. To enhance these benchmark datasets, web API crawlers were implemented to retrieve missing images and replies from the official Twitter and Weibo websites. This extension allows for an investigation and comparison of the importance of multimodal source and propagation graph features under unified benchmark datasets. The availability of these extended datasets can facilitate future research in the field of multimodal-graph rumour detection.

In real-world scenarios, social media conversations often contain missing data features, such as images or replies. To address this challenge, an attention mask is employed in the attention mechanism to handle missing modalities. The attention mask enables the fusion process to ignore missing modal information when generating meaningful representations of a conversation for rumour detection. Consequently, the proposed network can handle input data with incomplete modalities during both training and testing, making it particularly suitable for real-world applications. Generally, a Transformer network for rumour detection experiences a performance drop when dealing with data with missing features. To overcome this, a multitask learning framework is employed to train the network using data that includes both complete and missing features. Experimental results demonstrate that adopting the multitask learning framework can significantly enhance the network's robustness to input samples with missing features. With improved robustness to missing modalities, the proposed network can be deployed as a single model in real-world scenarios, effectively detecting rumours from data with complete or missing features.

## 6. Conclusion and Future Work

In conclusion, this paper proposes the integration of two common approaches, multimodality-based and propagation-based rumour detection, through unified learning models and shared datasets. The Unified Multimodal Graph Transformer Network (UMGTN) is introduced as a solution for modelling social media conversations and detecting rumours. UMGTN comprises the Multimodal Transformer Network (MTN), Graph Transformer Network (GTN), and

Multimodal Graph Transformer (MGT), which extract and fuse multimodal source and propagation graph information from conversations. Additionally, a multitask learning framework is leveraged to train the network using data with missing features, thereby enhancing its robustness and flexibility. Experimental results confirm the effectiveness of the proposed framework on four real-world benchmark datasets for rumour detection, encompassing English and Chinese conversations from Twitter and Weibo. The study also reveals that rumours spread on social media can be written in different languages. In future work, transfer learning will be employed for cross-lingual rumour detection to identify rumours written in low-resource languages with limited available samples.